\documentstyle[11pt,menu97]{article}

\begin{document}
    \setlength{\baselineskip}{2.6ex}

\title{Dispersion Theoretical Analysis of Pion Photoproduction\\
at Threshold and in the Delta Region}
\author{L.~Tiator\\
{\em Institut f\"ur Kernphysik, Universit\"at Mainz, J.J. Becherweg 45}}

\maketitle

\vspace{-2.5cm}
\hfill MKPH-T-97-29
\vspace{1.9cm}

\begin{abstract}
  \setlength{\baselineskip}{2.6ex} A new partial wave analysis of pion
  photoproduction has been obtained in the framework of fixed-$t$
  dispersion relations valid form threshold up to 500 $MeV$. It is
  based on new Mainz data for $\pi^{\circ}$ and $\pi^{+}$ production
  off the proton and both older and more recent data from Bonn,
  Frascati and TRIUMF for $\pi^{+}$ and $\pi^{-}$.  At threshold we
  obtain a good agreement with the existing data for both charged and
  neutral pion production. In the resonance region we have precisely
  determined the electromagnetic properties of the $\Delta(1232)$
  resonance, in particular the E2/M1 ratio $R_{EM}=-2.5 \pm 0.1 \%$.
  We show that a model independent resonance background subtraction is
  possible with the speed-plot technique and obtain the $\Delta$ pole
  at $W =(1211 - 50i)$ MeV and the E2/M1 ratio of the residues as
  $R_{\Delta} = - 0.035 - 0.046 i$.
\end{abstract}

\section*{INTRODUCTION}
Dispersion theoretical analyses have been very successful in the
description and understanding of pion nucleon scattering and pion
photoproduction already in the 60s. Based on unitarity, analyticity,
crossing symmetry, gauge invariance and Lorentz invariance they
provide a powerful tool to investigate the low energy behaviour of
the nucleon and the structure of nucleon resonances.  During the last
few years beams of high current and high duty factor together with
considerably improved particle detection techniques have reduced the
statistical errors to the order of a few percent, and promise to keep
control of the systematical errors at the same level. To interpret
these data with respect to the most interesting features, i.e. the
threshold behaviour and the electromagnetic excitation of resonances,
a partial wave analysis is mandatory. To ensure the consistency and
uniqueness of such an analysis, constraints from unitarity and
dispersion relations have to be imposed. Such concepts have proven to
be quite successful in pion-nucleon scattering \cite{Hoe83}. In
comparison with that field, the situation in pion photoproduction is
considerably more complex. The spin and isospin structure leads to
twelve independent amplitudes, while in pion-nucleon scattering there
are only four such amplitudes. As a consequence a complete experiment
requires the use of many polarization observables. Such a complete
experiment has not yet been performed. However, the new experiments
provide an ever increasing amount of precise and new data.  At
present, the experimental thrust is mainly on measurements near
threshold and around the $\Delta$(1232) resonance. In the coming
years, a series of experiments at Jefferson Lab will cover the whole
resonance region.  Restricting our theoretical investigations to the
threshold region and the low-lying resonances, we are lead to choose
the method of Omn\`es and Muskhelishvili to analyse the existing data,
because it introduces a natural parametrization and fulfills the
constraints of unitarity at the same level.

\section*{DISPERSION RELATIONS AT FIXED t}

Starting from fixed-$t$ dispersion relations for the invariant amplitudes
 of pion photoproduction, the projection of the multipole amplitudes leads
to a well known system of integral equations,
\begin{equation} \label{inteq}
\mbox{Re}{\cal M}_{l}(W) = {\cal M}_{l}^{\mbox{\scriptsize P}}(W)
+ \frac{1}{\pi}\sum_{l'}{\cal P}\int_{W_{\mbox{\scriptsize thr}}}^{\infty}
K_{ll'}(W,W')\mbox{Im}{\cal M}_{l'}(W')dW',
\end{equation}
where ${\cal M}_l$ stands for any of the multipoles
$E_{l\pm}, M_{l\pm},$ and ${\cal M}_{l}^{\mbox{\scriptsize P}}$ for the
corresponding (nucleon) pole
term. The kernels $K_{ll'}$ are known, and the real and imaginary parts of the
amplitudes are related by unitarity. In the energy region below two-pion
threshold, unitarity is expressed by the final state theorem of
Watson,
\begin{equation} \label{watson}
\cal{M}_{l}^{I} (W) = \mid \cal{M}_{l}^{I} (W)\mid e^{i(\delta_{l}^{I} (W)
+ n\pi)},
\end{equation}
where $\delta_{l}^{I}$ is the corresponding $\pi N$ phase shift and
$n$ an integer. We have essentially followed the method of Schwela et
al \cite{Sch69,Pfe72} to solve Eq. (\ref{inteq}) with the constraint
(\ref{watson}). In addition we have taken into account the coupling to
some higher states neglected in that earlier reference. At the
energies above two-pion threshold up to $W = 2$ GeV,
Eq.~(\ref{watson}) has been replaced by an ansatz based on unitarity
\cite{Sch69}. Finally, the contribution of the dispersive integrals
from $2$ GeV to infinity has been replaced by $t$-channel exchange,
parametrized by certain fractions of $\rho$- and $\omega$-exchange.
Furthermore, we have to allow for the addition of solutions of the
homogeneous equations to the coupled system of Eq.~(\ref{inteq}). The
whole procedure introduces 9 free parameters, which have to be
determined by a fit to the data.\cite{HDT}

In our data base we have included the recent MAMI experiments for
$\pi^{\circ}$ and $\pi^{+}$ production off the proton in the energy
range from 160 MeV to 420 MeV\cite{Fuc96,Bec96,Hae96}, both older and
more recent data from Bonn for $\pi^+$ production off the
proton\cite{Men77,Bue94,Dut95}, and older Frascati\cite{Car73} and
more recent TRIUMF data\cite{Bag88} on $\pi^{-}$ production off the
neutron. Our fit obtained with this data base describes this data very
well and in addition it also gives good agreement with data from the
world data base (e.g. SAID\cite{Arn96}) not included in our fit. 

\section*{RESULTS FOR THE THRESHOLD REGION}

In Table 1 we give our results for the $s$- and $p$-wave amplitudes at
threshold and compare them to the available information from heavy
baryon chiral perturbation theory (HBChPT) \cite{Ber96b} and
experiment.  The reduced $p$-wave amplitudes are defined as usual by
$m_1={\cal M}_1/(\mid\vec k\mid\mid\vec q\mid)$, in the limit $k
\rightarrow 0$. In fact the dependence on the photon energy $q$ is not
stringent, and in explicit calculations this definition leads to less
energy dependence for $e_{1+}$ and $m_{1+}$, while $m_{1-}$ would vary
less without the factor $\mid\vec q\mid$.  In general there is a very
good agreement between our analysis, HBChPT and experiment. In our
analysis the uncertainties are mainly in the neutron channel,
especially for $n(\gamma,\pi^0)n$, where the lack of precise
experimental data, in particular of polarization observables, reflects
in the threshold values.  In fact it is interesting to note that the
threshold amplitudes given in Table 1 are not fitted to the threshold
data.  They are only determined from experimental information above
$160$ MeV, therefore, they should be considered as predictions rather
than fits. The interplay of the complete knowledge of pion
photoproduction at all energies in the framework of dispersion
relations can be seen in the individual contributions to the threshold
values of the $s$-wave amplitudes. For $p(\gamma,\pi^0)p$ we obtain
\[
E_{0+}^{\mbox{thr}}=-7.63+4.15+2.32-0.41+0.29+0.07=-1.22,
\]
where the individual contributions are from the pole terms, from 
$M_{1+}, E_{0+}, E_{1+}, M_{1-}$ and higher multipoles, respectively.

There is special theoretical interest in the $E_{0+}$ amplitude of
$n(\gamma,\pi^-)p$ because it allows for an independent determination
of the charge exchange pion-nucleon scattering length via the Panofsky
ratio, $P = \sigma(\pi^{-}p\to\pi^{0}n)/\sigma(\pi^{-}p\to\gamma n)$.
This ratio is well determined by experiment, $P=1.543\pm 0.008$
\cite{Spu77}, and related to the scattering length by time reversal,
\begin{equation}
 \label{scat_mul}
  a_{CEX}\equiv a(\pi^-p\to\pi^0n)=\sqrt{2 \frac{q_0}{k_{0}} P}
  \; E_{0+}^{\mbox{thr}}(\pi^{-}p),
\end{equation}
where $q_0$ and $k_0$ are the $cm$ momenta of photon and neutral pion
at $p \pi^-$ threshold.  Using our value of the threshold amplitude,
and the measured Panofsky ratio, we find $a_{CEX} = (-0.120 \pm 0.002)\times
m_{\pi}^{-1}$.  This has to be compared with the value $(-0.129 \pm
0.002)\times m_{\pi}^{-1}$ resulting from a partial wave analysis of
pion-nucleon scattering \cite{Hoe83} (solution KH80). Recently,
$a_{CEX}$ has also been determined by studying the level spacing of
pionic atoms, with a preliminary value of $(-0.1301 \pm 0.0059)\times
m_{\pi}^{-1}$ \cite{Jan96}.
\begin{table}[htbp]
    \caption{Threshold amplitudes for pion photoproduction. 
      The $s$-waves $E_{0+}$ are in units of $10^{-3}/m_{\pi^{+}}$ and
      the reduced $p$-wave multipoles are in units of
      $10^{-3}/m_{\pi^{+}}^{3}$.  Our values are compared with results
      from chiral perturbation theory \protect\cite{Ber96b} and data
      analysis for charged pion production\protect\cite{Ada76} and 
      neutral pion production off the proton\protect\cite{Fuc96,Berg97}.}
  \begin{center}
    \leavevmode
    \begin{tabular}{|l|cccc|cccc|}
  \hline
     & \multicolumn{4}{|c|}{$\gamma p\to\pi^{+}n$} 
     & \multicolumn{4}{c|}{$\gamma n\to\pi^{-}p$} \\
  \hline
    & $E_{0+}$ & $m_{1-}$ & $e_{1+}$ &
      $m_{1+}$ & $E_{0+}$ & $m_{1-}$ &
      $e_{1+}$ & $m_{1+}$ \\
  \hline
    Disp. & 28.0$\pm$0.2 & 6.1 & 4.9 & -9.6 
          & -31.7$\pm$0.2 & -8.3 & -4.9 & 11.2 \\
    ChPT  & 28.2$\pm$0.6 & & & & -32.7$\pm$0.6 & & & \\
    Exp.  & 28.3$\pm$0.3 & & & & -31.8$\pm$1.9 & & & \\
  \hline\hline
     & \multicolumn{4}{|c|}{$\gamma p\to\pi^{0}p$} 
     & \multicolumn{4}{c|}{$\gamma n\to\pi^{0}n$} \\
  \hline
    & $E_{0+}$ & $m_{1-}$ & $e_{1+}$ &
      $m_{1+}$ & $E_{0+}$ & $m_{1-}$ &
      $e_{1+}$ & $m_{1+}$ \\
  \hline
    Disp. & -1.22$\pm$0.16 & -3.92 & -0.15 & 7.07 
          &  1.19$\pm$0.16 & -2.16 & -0.17 & 5.97 \\
    ChPT 
          & -1.16 & -3.21& -0.11& 7.45
          &  2.13 & -1.63& -0.16& 6.25\\
    Exp. 
          & -1.31$\pm$0.08 & -3.38$\pm$0.26& -0.25$\pm$0.17& 7.44$\pm$0.04
          & & & & \\
  \hline
    \end{tabular}
    \label{tab:s_pi_pm}
  \end{center}
\end{table}

\section*{RESULTS FOR THE RESONANCE REGION}

According to Watson theorem, at least up to the two-pion threshold,
the ratio $E_{1+}^{(3/2)}/M_{1+}^{(3/2)}$ is a real quantity. However,
it is not a constant but even a rather strongly energy dependent
function. If we determine the resonance position as the point, where
the phase $\delta_{1+}^{(3/2)}(W=M_\Delta)=90^\circ$, we can define
the so-called "full" ratio
\begin{equation}
R_{EM} = \left. \frac{E_{1+}^{(3/2)}}{M_{1+}^{(3/2)}}\right|_{W=M_\Delta}
       = \left. \frac{\mbox{Im}E_{1+}^{(3/2)}}{\mbox{Im}M_{1+}^{(3/2)}}
       \right|_{W=M_\Delta}\,.
\end{equation}
We note that this ratio is identical to the ratio obtained with the
$K$-matrix at the $K$-matrix pole $W=M_\Delta$. This can be seen by
using the relation between the $T$- and the $K$-matrix, $T=K
\cos\delta e^{i\delta}$ and consequently $K=\mbox{Re}T+\mbox{Im}T\,
tan\delta$. Therefore, at $W=M_\Delta$ we find
$K(E_{1+}^{(3/2)})/K(M_{1+}^{(3/2)})=
\mbox{Im}E_{1+}^{(3/2)}/\mbox{Im}M_{1+}^{(3/2)}=R_{EM}$.  The recent,
nearly model-independent value of the Mainz group at $W
=M_{\Delta}=1232$ MeV is $(-2.5 \pm 0.2 \pm 0.2)\%$ \cite{Bec96} is in
excellent agreement with our dispersion theoretical calculation that
gives $(-2.5\pm 0.1)\%$.

The analytic continuation of a resonant partial wave as function of
energy into the second Riemann sheet should generally lead to a pole
in the lower half-plane. A pronounced narrow peak reflects a
time-delay in the scattering process due to the existence of an
unstable excited state. This time-delay is related to the speed $SP$
of the scattering amplitude $T$, defined by\cite{Hoe92}
\begin{equation}
SP(W) = \left\vert \frac{dT(W)}{dW}\right\vert ,
\end{equation}
where $W$ is the total $c.m.$ energy. In the vicinity of the resonance
pole, the energy dependence of the full amplitude $T = T_{B} + T_{R}$
is determined by the resonance contribution,
\begin{equation}\label{T_res}
  T_{R} (W) = \frac{r\Gamma_{R} e^{i\phi}}{M_{R}- W-
    i\Gamma_{R}/2}\,\,,
\end{equation}
while the background contribution $T_B$ should be a smooth function of
energy, ideally a constant. We note in particular that $W_R = M_{R}-
i\Gamma_{R}/2$ indicates the position of the resonance pole in the
complex plane, i.e. $M_{R}$ and $\Gamma_{R}$ are constants and differ
from the energy-dependent widths, and possibly masses, derived from
fitting certain resonance shapes to the data.

Applying this technique to our $P_{33}$ amplitudes we find the pole at
$W_{R} = M_{R} - i \Gamma_{R}/2 = (1211 - 50 i)$ MeV in excellent
agreement with the results obtained from $\pi N$ scattering, $M_{R}=
(1210\pm 1)$ MeV and $\Gamma_{R} = 100$ MeV \cite{Hoe92}.  The complex
residues and the phases are obtained as $r_E=1.23\cdot
10^{-3}/m_{\pi}, \phi_E=-154.7^{\circ}, r_M=21.16\cdot
10^{-3}/m_{\pi}$ and $\phi_M=-27.5^{\circ}$, yielding a complex ratio
of the residues
\begin{equation}
R_{\Delta} = \frac{r_{E} e^{i\phi_{E}}}{r_{M} e^{i\phi_{M}}}
= - 0.035 - 0.046 i.
\end{equation}
While the experimentally observed ratio $R_{EM}$ is real and very
sensitive to small changes in energy, the ratio $R_{\Delta}$ is a
complex number defined by the residues at the pole, therefore, it does
not depend on energy.

It should be noted, however, that a resonance without the
accompanying background terms is unphysical, in the sense that only
the sum of the two obeys unitarity. Furthermore we want to point out
that the speed-plot technique does not give information about the
strength parameters of a "bare" resonance, i.e. in the case where the
coupling to the continuum is turned off. Both the pole position and
the residues at the pole will change for such a hypothetical case, but
the exact values for the "bare" resonance can only be determined by a
model calculation and as such will depend on the ingredients of the
model.
\begin{table}[htbp]
\protect
    \caption{E/M ratios of different analyses. $R_{EM}$ gives the
    "full" ratio $E_{1+}^{(3/2)}/M_{1+}^{(3/2)}$ at
    $W=M_\Delta$ and $R_\Delta$ gives the complex ratio obtained by the
    speed-plot technique at the resonance pole. All numbers are given
    in percentage.}
\small
\vspace{0.4cm}
\begin{center}
\begin{tabular}{|l|c|c|}
\hline
    analysis & $R_{EM}$ [\%] & $R_\Delta$ [\%]\\
             & at $W_\Delta=1232$ MeV & at $W_{R}=(1211- 50 i)$ MeV\\  
\hline
    VPI (SP97)\protect\cite{Arn96} & -1.4 & -3.1 -5.0 i  \\
    VPI (B500)\protect\cite{VPI97a} & -2.5 & -4.0 -3.5 i  \\
\hline
    RPI \protect\cite{Dav90} & -3.19 & -4.8 -4.6 i  \\
\hline
    this work                & -2.54 & -3.5 -4.6 i  \\  
\hline
    Mainz experiment\protect\cite{Bec96}
                             & $-2.5 \pm 0.2 \pm 0.2 $ & \\  
\hline
    BNL experiment\protect\cite{Bla97}
                             & $-3.0 \pm 0.3 \pm 0.2 $ & \\  
\hline
    \end{tabular}
    \label{tab:ratios}
  \end{center}
\end{table}
In Table 2 we compare our result with two VPI solutions, the solution
of the RPI group in a field theoretical Lagrangian approach and the
experimental analyses of Mainz and Brookhaven.  With the numerical
solutions of VPI and RPI we have applied the speed-plot technique in
order to separate resonance and background contributions and to
determine the pole position and residues. While the "full" ratios
$R_{EM}$ vary by more than a factor of 2 among these solutions, the
ratios $R_\Delta$ are much closer to each others. In particular, the
imaginary parts are very stable within only about 30\%.

Finally, we have determined the photon couplings $A_{1/2}$ and
$A_{3/2}$ of the delta resonance.
From our energy-dependent analysis we get $A_{1/2}=(-132\pm2)$
and $A_{3/2}=(-253\pm3)$, both in units of $10^{-3}/\sqrt{GeV}$.
However, it should be again noted that both the $R_{EM}$ ratio and the
photon couplings, calculated at the K-matrix pole are well-defined
quantities but they have no direct connection to quark model
calculations of a "bare" resonance.

\section*{SUMMARY}
With the new and very precise data obtained at MAMI in Mainz we have
obtained a new partial wave analysis for pion photoproduction. The
uncertainties in most multipoles could be considerably improved
compared to previous analyses. Very accurate results can be obtained
at threshold and in the resonance region.  At resonance we must
clearly distinguish between the resonance position $W_{\Delta}=1232$
MeV on the real axis and the pole at $W_R =(1211 - 50i)$ MeV in the
complex plane.  At the resonance position, where the phase passes
$90^{\circ}$, we obtain an REM ratio of $R_{EM}=(-2.5 \pm 0.1) \%$ in
very good agreement with the experimental analysis\cite{Bec96}. This
was also recently confirmed in a VPI analysis with a restricted data
base\cite{VPI97a}.  At the pole in the complex plane we obtain the
ratio of the resonant electric and magnetic multipoles as $R_{\Delta}
= - 0.035 - 0.046 i$.  This is a model-independent ratio that can be
determined in any analysis or calculation of pion photoproduction.
After a long time of confusion about the different ratios that can be
defined and constructed out of the measured cross sections or the
analysed multipoles $E_{1+}^{(3/2)}$ and $M_{1+}^{(3/2)}$, it now
appears that the ratio $R_\Delta$ is the closest one can get to a
background subtracted value.  Such a ratio must be complex, and it
will be a challenge for all microscopic models to determine this
ratio.

This work was supported by the Deutsche Forschungsgemeinschaft (SFB~201).
             

\bibliographystyle{unsrt}

\end{document}